\documentclass[12pt,preprint]{aastex}

\begin{document}
\title{Origin of the X-rays and Possible GeV-TeV Emission from the Western Hot Spot Of Pictor A}
\author{
Jin Zhang\altaffilmark{1,2,3}, J. M. Bai\altaffilmark{1}, Liang
Chen\altaffilmark{1,3}, Xian Yang\altaffilmark{1}}
\altaffiltext{1}{National Astronomical Observatories/Yunnan
Observatory, Chinese Academy of Sciences, P. O. Box 110, Kunming,
Yunnan, 650011, China;}\altaffiltext{2}{College of Physics and
Electronic Engineering, Guangxi Teachers Education University,
Nanning, Guangxi, 530001, China;
\\}\altaffiltext{3}{The Graduate School of Chinese Academy of
Sciences\\}

\begin{abstract}
Pictor A is a nearby Fanaroff-Riley class II (FR II) radio galaxy with a bright hot spot, the western hot spot.
Observation of high polarization in the optical emission of the hot spot indicates that the optical emission could be
synchrotron radiation of relativistic electrons in the hot spot. These electrons may be able to produce high energy
$\gamma$-ray photons through inverse Compton (IC) scattering. We use single-zone and multi-zone synchrotron +
synchrotron-self-Compton (SSC) models to fit the observed spectral energy distribution (SED) from the radio to the
X-ray band of the hot spot. Our results show that in the case of a much weaker magnetic field strength than the
equipartition magnetic field, both the single-zone and multi-zone models can fit the SED, but the multi-zone model
significantly improves the fit. The two models predict the hot spot as a GeV-TeV source, which might be marginally
detectable with \emph{Fermi}/LAT and HESS. The inverse Compton scattering of cosmic microwave background (IC/CMB) is
also considered, but its contribution to GeV-TeV emission is negligible. Note that under the equipartition condition,
the SED can also be fit with the multi-zone model, but the predicted flux at $>10^{22}$ Hz is too weak to be
detectable. The detection of TeV $\gamma$-rays from this FR II radio galaxy, if confirmed, would establish a new
subclass of extragalactic source in this energy regime since most of the AGNs detected to date at TeV energies are
high-energy-peaked BL Lac objects.
\end{abstract}



\section{Introduction}           
\label{sect:intro} Blazars, which include both BL Lac objects and
flat spectrum radio quasars (FSRQs), are typically characterized
by a relativistic jet with a small angle between the jet axis and
the line of sight. Their radiations are dominated by the jet. In
the unified schemes (Urry \& Padovani 1995), FR I radio galaxies
are the parent population of BL Lac objects, and FR II radio
galaxies are of FSRQs. Up to now, more than 20 blazars are
identified as TeV sources (De Angelis et al. 2008; Aharonian et
al. 2008a). Most of them are X-ray bright, high-energy-peaked BL
Lac (HBL) objects, with exceptions of one low-energy-peaked BL Lac
object (LBL/BL Lacertae, Albert et al. 2007), one FSRQ (3C 279,
Albert et al. 2008), two intermediate-energy-peaked BL Lac Objects
(IBL/W Comae, Acciari et al. 2008; IBL/3C 66A, Acciari et al.
2009), and one blazar with uncertain classification (S5 0716+71;
Teshima 2008). Interestingly, Bai \& Lee (2001) predicted that M87
and Centaurus A, two FR I radio galaxies, are TeV emission
sources, and they were confirmed with the observations of HESS
(Aharonian et al. 2003; Aharonian et al. 2009).

Hot spots are mostly found at the end of the lobes in FR II radio galaxies. They are thought to be the location of jet
termination and are characterized  by both high surface brightness and prominent location near the outmost boundaries
of radio lobes (Fnanaroff \& Riley 1974; Blandford \& Rees 1974; Begelman et al. 1984; Bicknell 1985; MeisenHeimer et
al. 1989). Hot spots are believed to be the accelerated sites of relativistic electrons. Furthermore, the high
polarization observation of some optical hot spots indicates that the optical emission is from synchrotron radiation of
relativistic electrons in the hot spots (Roeser \& Meisenheimer 1987; L\"{a}hteenm\"{a}ki \& Valtaoja 1999). Assuming a
magnetic field strength $B \sim 10^{-5}$ G, one can estimate the energy of relativistic electrons $\gamma \sim 10^{6}$,
which contribute the optical emission by synchrotron process. If the observed X-rays also come from the synchrotron
process, the energy of relativistic electrons is even higher. These electrons may interact with the synchrotron photons
to produce very high energy $\gamma$-ray photons by inverse Compton (IC) scattering. This motivates us to investigate
the X-ray emission from the hot spots and to examine whether their very high energy (VHE) emission can be detectable
with HESS and \emph{Fermi}.

The origin of the X-ray emission is highly debated. The observed
SEDs from the radio to X-ray band of some hot spots are explained
as a single synchrotron radiation (Harris et al. 1998; Worrall et
al. 2001; Hardcastle et al. 2004). However, an IC component is
required to model the X-ray emission for most of the hot spots
(e.g., Zhang et al. 2009). Although the SSC model can be used to
explain the X-ray emission of some hot spots, the equipartition
condition would be enormously violated (Hardcastle et al. 2002,
2004; Kataoka et al. 2003; Kino \& Takahara 2004; Tavecchio et al.
2005). It was also proposed that the X-ray emission may be
contributed by an electron population different from that for the
radio-optical emission (Harris et al. 2004), and that the
relativistic beaming effect in the hot spots may also be taken
into account (Georganopoulos \& Kazanas 2003; Tavecchio et al.
2005).

Pictor A is the prototype of FR II radio galaxy at a redshift
$z=0.035$ with a remarkable primary hot spot, the western hot spot
(WHS). It was first discovered and named by Stanley \& Slee
(1950). Its basic double structure was reported by Maltby \&
Moffet (1962), and some hot spots located at the lobes were
resolved by Schwarz et al. (1974). The WHS locates at 4.2 arcmin
from the nucleus. It is much brighter than the eastern one, and is
a remarkable object with its high radio brightness and the
observation of optical polarization (Roeser \& Meisenheimer 1987;
Thomson et al. 1995) among the brightest radio hot spots in radio
galaxies and quasars (Wilson et al. 2001). Perley et al. (1997)
resolved the WHS in much higher resolution with the observation by
the Very Large Array (VLA). The observations in the infrared band
were reported by Meisenheimer et al. (1997). The X-ray
observations of the WHS were first made with the Einstein
telescope (Roeser \& Meisenheimer 1987), and then significantly
improved by the Chandra X-ray telescope, which derived an X-ray
spectral index $\alpha=1.07\pm0.11$ and a luminosity
$1.7\times10^{42}$ ergs $\rm{s}^{-1}$ in the 2-10 keV band  in
2000 January (Wilson et al. 2001). Aharonian et al. (2008b)
observed Pictor A with the HESS telescope, but no significant
detection was obtained. The high luminosity and observed broadband
SED from radio to X-ray band make the WHS an important candidate
for studying the high-energy processes in the hot spots.

Interestingly, Tingay et al. (2008) resolved the WHS into some
bright sub-components embedded in a diffuse emission region with
Very Long Baseline Array (VLBA), and suggested that the X-ray
emission of the WHS may come from the synchrotron process of some
compact sub-components.  In this paper, we focus on the physical
origin of the X-rays and possible high energy gamma-rays from the
hot spot. The sychrotron + SSC model for two cases that the
radiations from the radio to X-ray band are from the same region
(single-zone model) and the X-rays are from a region different
from that of the radio/optical emission (multi-zone model) are
used to fit the observed SED. Our model is described in \S2, and
the results are reported in \S 3. Conclusions and discussion are
present in \S 4. Throughout, a concordance cosmology with
parameters $H_0 = 70$ km s$^{-1}$ Mpc$^{-1}$, $\Omega_M=0.30$, and
$\Omega_{\Lambda}=0.70$ are adopted.

\section{Model}
\label{sect:Model} The observed SEDs of blazars are characterized with a double-peaked structure, generally being
responsible for the synchrotron emission of the relativistic electrons and the inverse Compton scattering of the
synchrotron photons by the same electron population (SSC; Maraschi et al. 1992; Bloom \& Marscher 1996). Although the
double-peaked structure of the SED for the WHS is not well-established, some signatures were observed. The observation
of polarization indicates that the radio and optical fluxes of the WHS should be produced by synchrotron radiation
(Roeser \& Meisenheimer 1987; Thomson et al. 1995; Perley et al. 1997), and the SED from the radio to the X-ray band
shows that the X-rays are not a simple extension of the radio and the optical emission. We fit the broadband SED of the
WHS with single- and multi-zone synchrotron + SSC models.

We assume that the emitting region is a homogeneous sphere with
radius $R$, and the electron distribution as a function of energy
($\gamma$) is assumed to be a broken power law,
\begin{equation}
N(\gamma )=\left\{ \begin{array}{ll}
                    N_{0}\gamma ^{-p_1}  &  \mbox{ $\gamma \leq \gamma _b$}, \\
            N_{0}\gamma _b^{p_2-p_1} \gamma ^{-p_2}  &  \mbox{ $\gamma > \gamma _b$,}
           \end{array}
       \right.
\end{equation}
where $p_{1}=2\alpha_{1}+1$ and $p_{2}=2\alpha_{2}+1$ are the indices below and above the break energy
$\gamma_{b}m_{e}c^{2}$, respectively, and $\alpha_{1,2}$ are the observed spectral indices. The $\gamma_{b}$ is
determined by peak frequency of the synchrotron radiation
\begin{equation}
\nu_{s}=\frac{4}{3}\nu_B\gamma_{b}^{2}\frac{\delta}{1+z},
\end{equation}
where $\nu_{B}=2.8\times10^6 B$ Hz is the Larmor frequency of
electrons in magnetic field $B$ and $\delta$ is the Doppler
factor. Although the detection of one-sidedness at X-rays may
indicate that the relativistic beaming effect should play a role
(Tavecchio et al. 2005) and a mean outward velocity of
$0.11\pm0.013c$ was also obtained for some hot spots (Arshakian \&
Longair 2000), the relativistic beaming effect is still quite
uncertain. Thus we do not consider this effect for the WHS in our
model and take $\delta=1$.

The total luminosity of the synchrotron emission in the comoving
frame is calculated by
\begin{equation}
L^{'}_{syn}=\int_{\gamma_{min}}^{\gamma_{max}}
N(\gamma)\bar{P}(\gamma)V d \gamma,
\end{equation}
where $V=4\pi R^{3}/3$ is the volume of radiation region and
$\bar{P}(\gamma)$ is the single electron synchrotron emission
power averaged over an isotropic distribution of pitch angles. It
is given by
\begin{equation}
\bar{P}(\gamma)=\frac{4}{3}\sigma_{T}c\beta^{2}\gamma^{2}U_{B}.
\end{equation}
Here $\beta=v/c$ can approximately be taken for 1,
$\sigma_{T}=8\pi e^{4}/(3m_{e}^{2}c^{4})$ is the Thomson cross
section, and $U_{B}=B^{2}/8\pi$ is the magnetic energy density.
According the equations mentioned above, we can obtain the
electron density parameter $N_{0}$ by
\begin{equation}
N_{0}=\frac{2L_{syn}\gamma^{(p_{1}-3)}_{b}}{\frac{4}{3}\sigma_{T}c\beta^{2}U_{B}f(\alpha_{1},\alpha_{2})V\delta^{4}},
\end{equation}
where$L_{syn}=L^{'}_{syn}\delta^{4}$ is the observed total
luminosity of synchrotron emission, and $f(\alpha_{1},\alpha_{2})$
is given by
\begin{equation}
f(\alpha_{1},\alpha_{2})=\frac{1}{1-\alpha_{1}}+\frac{1}{\alpha_{2}-1}.
\end{equation}
The observed total luminosity of the synchrotron emission is derived from
\begin{equation}
L_{syn}=f(\alpha_{1},\alpha_{2})\nu_{s}L_{s}(\nu_{s}),
\end{equation}
where $L_{s}(\nu_{s})$ is the peak luminosity of the synchrotron
radiation at the peak frequency $\nu_{s}$.

The equipartition magnetic field is calculated with the formulae
given in Brunetti et al. (1997). With those approximations, the
synchrotron + SSC model can be completely specified by seven
parameters (Tavecchio et al. 1998), the magnetic field intensity
$B$, the emission region radius $R$, the Doppler factor $\delta$,
the slopes $p_{1}$ and $p_{2}$, the Lorentz factor of the
electrons at the break $\gamma_{b}$, and the electron density
parameter $N_{0}$. In our calculations, the the synchrotron
self-absorption to the low-frequency part of the spectrum is also
taken into account. In the GeV-TeV regime, the Klein- Nishina
effect could be significant. We consider this effect using the
following approximation (Zdziarski 1986),
\begin{equation}
\sigma=\left\{ \begin{array}{ll}
                    \sigma_{T}  &   \mbox{ $for \qquad \gamma x\leq \frac{3}{4}$} \\
            0  &   \mbox{ $for \qquad \gamma x> \frac{3}{4}$,}
           \end{array}
       \right.
\end{equation}
where $x=h\nu/m_{e}c^{2}$. This effect makes the IC spectrum show
a high-energy cut-off.

The observed flux derived from the model can be calculated using
the same numerical method given by Chiaberge \& Ghisellini (1999).
The frequency of the synchrotron radiation can be calculated by
equation (2). The synchrotron emissivity $\epsilon_s(\nu)$ is
given by
\begin{equation}
\epsilon_s(\nu)=\frac{1}{4\pi}\int_{\gamma_{min}}^{\gamma_{max}}d
\gamma N(\gamma)P_s(\nu, \gamma),
\end{equation}
where $P_s(\nu, \gamma)$ (see, e.g., Crusius \& Schlickeiser 1986;
Ghisellini et al. 1988) is the single electron synchrotron
emissivity averaged over an isotropic distribution of pitch
angles. According the transfer equation, the synchrotron radiation
field $I_s(\nu)$ is derived by
\begin{equation}
I_s(\nu)=\frac{\epsilon_s(\nu)}{k(\nu)}[1-e^{-k(\nu)R}],
\end{equation}
where $k(\nu)$ is the absorption coefficient (e.g. Ghisellini \&
Svensson 1991).

In the SSC scenario, the IC emissivity is calculated by
\begin{equation}\label{ec}
\epsilon_c(\nu_c)=\frac{\sigma_T}{4}\int_{\nu_i^{min}}^{\nu_i^{max}}
\frac{d\nu_i}{\nu_i}\int_{\gamma_1}^{\gamma_2}\frac{d\gamma}{\gamma^2\beta^2}N(\gamma)f(\nu_i,\nu_c)\frac{\nu_c}{\nu_i}I_s(\nu_i),
\end{equation}
where $\nu_i$ is the frequency of the incident photons emitted by
the synchrotron radiation between $\nu_i^{min}$ and $\nu_i^{max}$,
$\gamma_1$ and $\gamma_2$ are the lower and upper limits of the
scattering electrons, and $f(\nu_i,\nu_c)$ is the spectrum
produced by scattering monochromatic photons of frequency $\nu_i$
with a single electron (e.g. Rybicki \& Lightman 1979). The medium
is transparent for the IC radiation field, so we simply derived
$I_c(\nu_c)=\epsilon_c(\nu_c)R$. Assuming that $I_{s,c}$ is an
isotropic radiation field, then the observed flux density is given
by
\begin{equation}
F(\nu_{obs})=\frac{4\pi^2 R^2 I_{s,c}(\nu)\delta^3(1+z)}{4\pi
D^{2}},
\end{equation}
where \emph{D} is the luminosity distance of the source and
$\nu_{obs}=\nu\delta/(1+z)$.

\section{Results}
\label{sect:SED} The observed SED of the WHS is shown in Figures 1 and 2. The data are taken from Meisenheimer et al.
(1997), Wilson et al. (2001), and Tingay et al. (2008). The upper limit made with HESS, $2.45\times 10^{-12}$ photons
cm$^{-2}$ s$^{-1}$ above 320 GeV from Aharonian et al. (2008b), is also presented in the Figures.

\subsection{Single-zone synchrotron + SSC model}
The Hubble Space Telescope resolved the WHS with a diameter of 1.4 arcsec and the brightest region is $\sim 0.3$ arcsec
(FWHM, Thomson et al. 1995). Therefore the radiation region radius of the WHS is taken as $R=209$ pc (0.3 arcsec),
which is slightly smaller than that given in Wilson et al. (2001), $R=250$ pc. As shown in Figure 1, the
radio-optical-ultraviolet SED shows as a perfect synchrotron peak, but the X-ray emission neither is the simple
extension of the synchrotron radiation nor the IC component. We consider that the X-ray emission is contributed by both
the synchrotron radiation and the SSC component. In order to match the observed X-ray spectrum, the fit with this model
requires a magnetic field that is smaller than the equipartition magnetic field strength. Our good fit is shown in
Figure 1 and the fitting parameters are $\nu_{s}= 4.32\times10^{13}$ Hz, $\alpha_{1}=0.69$, $\alpha_{2}=1.45$ and
$B=4.4\times10^{-5}$ G. The model predicts a VHE emission in GeV-TeV band, which is even over the observed upper limit
given by Aharonian et al. (2008b). However, we should note that the fit cannot well describe the ultraviolet data.

\subsection{Multi-zone synchrotron + SSC model}
Based on the observation by Tingay et al. (2008) with VLBA, the
WHS is resolved into some bright sub-components embedded in a
diffuse emission region. We thus consider a multi-zone synchrotron
+ SSC model.

According to the observations in the radio (Perley et al. 1997; Tingay et al. 2008), optical (Thomson et al. 1995), and
X-ray (Wilson et al. 2001) bands, the radiation region is thought to contain a diffuse region of $R=975$ pc (1.4
arcsec) and five compact sub-components. We assume that the compact sub-components have the same luminosity and size
$R=60$ pc (average value of their sizes resolved by VLBA). We first fit the SED with a multi-zone model under the
equipartition condition. We take the same magnetic field strength as the average value of the equipartition magnetic
field $B_{eq}=4.0\times10^{-4}$ G for both the diffuse region and the sub-components. Our fit derives $\nu_{s}=
7.32\times10^{13}$ Hz, $\alpha_{1}=0.71$, $\alpha_{2}=1.72$ for the diffuse region and $\nu_{s}= 2.42\times10^{17}$ Hz,
$\alpha_{1}=0.75$, $\alpha_{2}=1.3$ for the sub-components. The result is shown in Figure 2. The observed SED is well
fit by this model. In this scenario, the emission in the radio-optical-ultraviolet band is produced by the relativistic
electrons through synchrotron process in the diffuse region, and the X-ray emission mainly comes from the synchrotron
radiation of the sub-components. The IC component in the GeV-TeV band predicted by this model is very weak, which is
much lower than the sensitivities of HESS and \emph{Fermi}/LAT.

The assumption of the energy equipartition condition has seldom
been possible to be tested, and there is no a strong priori reason
why it should be true (Hardcastle et al. 1998). Migliori et al.
(2007) suggested that the X-ray emission of the lobes in Pictor A
is due to IC/CMB and derived $B_{eq}/B_{IC} \sim 3$. Fan et al.
(2008) argued that the ratio should be higher in hot spots than
lobes. They suggested that the relativistic electrons  not only
experience radiative loss but also are subject to the adiabatic
expansion loss from hot spots to radio lobes same as the magnetic
field. Therefore, we take a small value of $B=1.5\times10^{-5}$ G,
under which the predicted flux should not exceed the upper limit
derived with HESS, and try to fit the SED with the multi-zone
model by keeping the other parameters not change. The result is
also shown in Figure 2. In this case, the model predicts a VHE
emission component in GeV-TeV band (dashed line in Figure 2),
similar to the single-zone model.

\subsection{Comparison of the SEDs between the WHS and the nucleus}
Since the space resolutions of \emph{Fermi} and HESS are
inadequate to resolve the WHS region
, one prominent issue regarding the GeV-TeV emission is that these high energy photons are from the nucleus of the
galaxy or from the WHS. Therefore, we fit the SED of the nucleus with the single-zone synchrotron + IC (SSC + EC) model
and then compare the result with the WHS. In this scenario, we also consider the contribution of IC scattering of
photon field of broad line region (EC/BLR) though it is very weak. The total luminosity of BLR is estimated by the
luminosities of the broad emission lines H$\alpha$ and H$\beta$ (Sulentic et al. 1995) according the equation (1) given
in Celotti et al. (1997). The BLR size is calculated with the formula (23) given in Liu \& Bai (2006). The radiation of
BLR is assumed to be a black body spectrum.

The observed SED of nucleus is taken from Perley et al. (1997) and Singh et al. (1990). The observations of Very Long
Baseline Interferometry (VLBI) indicate that the parsec-scale jet components of Pictor A are likely sub-luminal motion
(Tingay et al. 2000), so the beaming effect is considered for fitting the SED of the nucleus. Tingay et al. (2000)
suggested that the jet angle $\theta$ of Pictor A to our line of sight is less than $51^{\circ}$ within 10 micro arcsec
of the core. We take apparent velocity $\beta_{app}=1.1$ (the average value of three components in Tingay et al. 2000)
and $\theta=40^{\circ}$, and obtain the beaming effect factor $\delta=1.6$. The emission region radius is taken as
$R=10^{16}$ cm. The model fit derives $\nu_{s}= 4.0\times10^{13}$ Hz, $\alpha_{1}=0.18$, $\alpha_{2}=2.2$, and $B=7$ G.
The fitting result is shown in Figure 3 with comparison to that of the WHS. The synchrotron self-absorption is
significant for the nucleus and the radio emission should come from more extended region. As shown in Figure 3, the
high energy emission above 1 GeV is dominated by the WHS, indicating that the WHS should be a GeV-TeV source, if the
detection is confirmed.

\section{Conclusions and Discussion}
\label{sect:discussion}We have fit the observed SED of the WHS
with the single- and multi-zone synchrotron + SSC models. Our
results show that the X-ray emission of the WHS may be contributed
by both the synchrotron and the SSC radiations in the single-zone
model, but it may be only the synchrotron radiation of
relativistic electrons in the compact sub-components of the WHS in
the multi-zone model. Comparing the fits of the two models, the
multi-zone model greatly improves the fit, and the single-zone
model needs a magnetic field strength much smaller than the
equipartition field strength in order to fit the observed X-ray
spectrum. The single-zone model also cannot describe the
ultraviolet data well. The multi-zone model may be more reasonable
than the single-zone model as Tingay et al. (2008) reported.

Georganopoulos \& Kazanas (2003) reported that the bulk flow of
the material in radio galaxy hot spots may be mildly relativistic.
The energy density of the CMB in the jet frame is
$U_{CMB}=4\times10^{-13}(1+z)^{4}\Gamma^{2}$ ergs cm$^{-3}$, which
rapidly increases with the bulk Lorentz factor $\Gamma$ and the
redshift $z$ of the sources. The synchrotron radiation energy
density is $U^{'}_{syn}=L_{syn}/(4\pi R^{2}c\delta^{4})$ ergs
cm$^{-3}$. We examine whether the contribution of the IC
scattering of CMB photons by the relativistic electrons (IC/CMB)
becomes significant comparing with the SSC. In the single-zone
model, we get $U_{CMB}=4.6\times10^{-13}$ ergs cm$^{-3}$, which is
smaller than $U^{'}_{syn}=6.8\times10^{-11}$ ergs cm$^{-3}$ with
two orders of magnitude. Therefore, the contribution of IC/CMB to
VHE emission can be neglected comparing with the SSC, as shown in
Figure 1. In the multi-zone model, we have
$U^{'}_{syn}=1.9\times10^{-12}$ ergs cm$^{-3}$ for the diffuse
region and $U^{'}_{syn}=3.1\times10^{-11}$ ergs cm$^{-3}$ for the
sub-components, which are much higher than $U_{CMB}$.

Both the single- and the multi-zone models predict a VHE emission
component in the GeV-TeV band when the magnetic field strength is
smaller than the equipartition magnetic field with  an order of
magnitude. The predicted VHE emission is above the sensitivities
of \emph{Fermi} and HESS. Furthermore, we show that the GeV-TeV
emission from Pictor A is dominated by the WHS, but not the
nucleus, suggesting that the WHS should be a GeV-TeV emitter, if
the GeV-TeV detection is confirmed. The detection of the TeV
emission from Pictor A would establish a new subclass of
extragalactic source in this energy regime since most of the AGNs
detected to date at TeV energies are high-energy-peaked BL Lac
(HBL) objects. The predicted GeV-TeV flux by the models requires a
smaller magnetic field strength than the equipartition magnetic
field. The GeV-TeV observations might also be used to test the
equipartition condition for this source.

We notice that Pictor A was observed with the HESS TeV $\gamma$-ray telescope between January 2005 and July 2007, with
7.9 hours live time (Aharonian et al. 2008b), but no any significant TeV emission was detected during that observation.
This may be due to the sensitivity of HESS decreasing at larger zenith angle (Aharonian et al. 2008b). As shown in
Figures 1 and 2, the predicted GeV-TeV flux of the WHS is slightly over the sensitivities of \emph{Fermi} and HESS.
Longer exposure time may be needed in order to catch the TeV emission for the large zenith angle and the low fluxes of
the WHS. We should note that under the equipartition condition, the SED can also be fit with the multi-zone model. In
this scenario, the predicted flux in the range of  $>10^{22}$ Hz is too weak to be detected.

Pictor A is the list of 206 ``Very Important AGN/blazars" targets of Fermi for simultaneous multi-frequency analysis
campaign\footnote{http://glastweb.pg.infn.it/blazar/}. The detection of the high energy emission and the broadband SEDs
observed quasi-simultaneously would place much stronger constraints on the radiation mechanism and on the physical
parameters of the source.

\acknowledgments

This work was supported by the National Natural Science Foundation
of China under grants 10533050 and the National Basic Research
Program (¡±973¡± Program) of China under Grant 2009CB824800. J. M.
B. thanks support of the Bai Ren Ji Hua project of the CAS.

\clearpage
\begin{figure}
\plotone{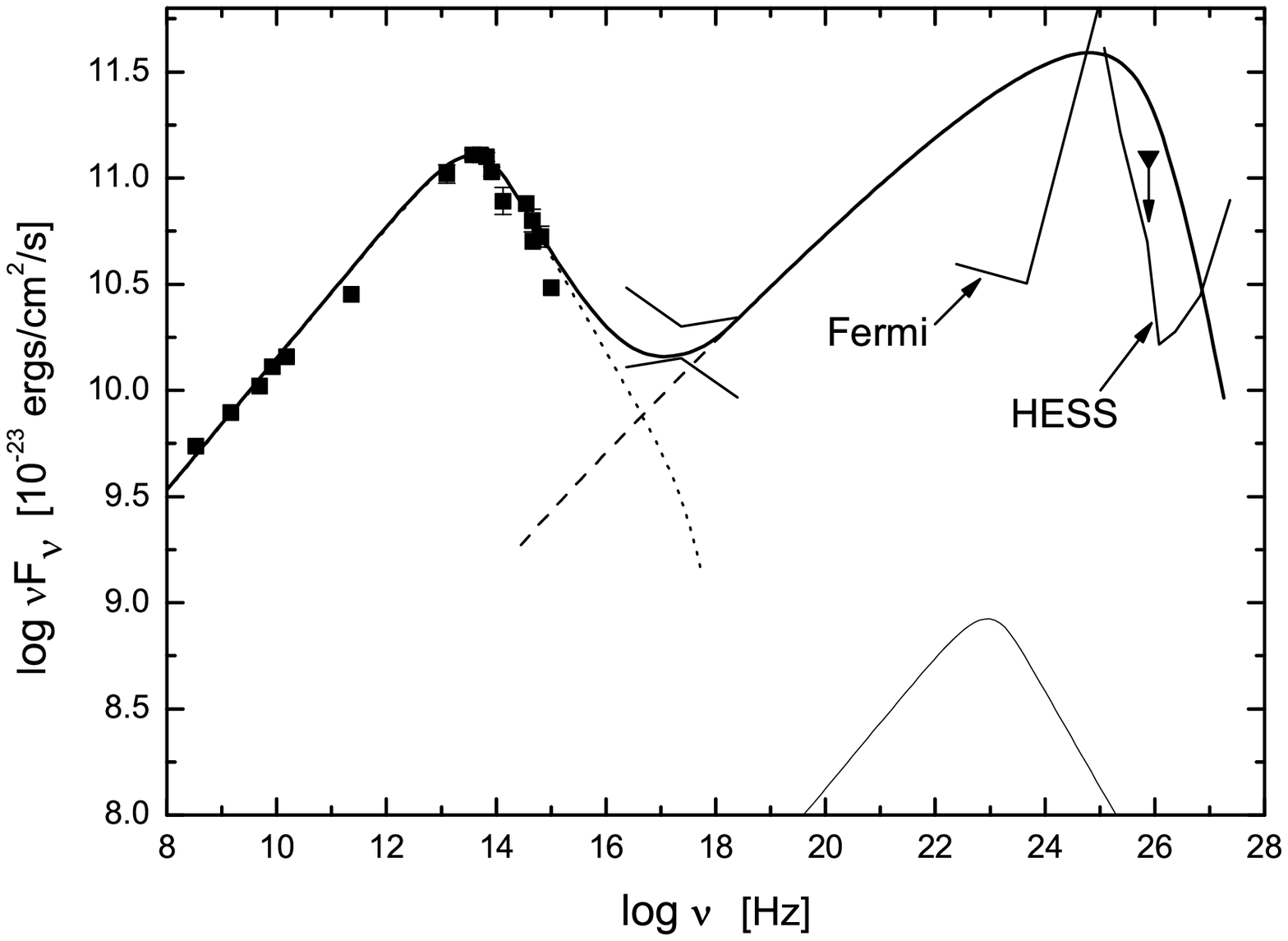} \caption{Broadband spectrum, plotted as $\nu
F_{\nu}$, of the western hot spot of Pictor A from the radio to
X-ray band. The radio and optical data are taken from Meisenheimer
et al. (1997). The infrared data are given by Tingay et al.
(2008). The two optical/near-ultraviolet points and the "bow tie"
of X-ray spectrum are from Wilson et al. (2001). The VHE upper
limit is from Aharonian et al. (2008b). The thick solid line is
the sum of synchrotron emission (dotted line) and a SSC component
(dashed line). The thin solid line is the spectrum of IC/CMB.
 \label{Fig:1}}
\end{figure}

\clearpage
\begin{figure}
\plotone{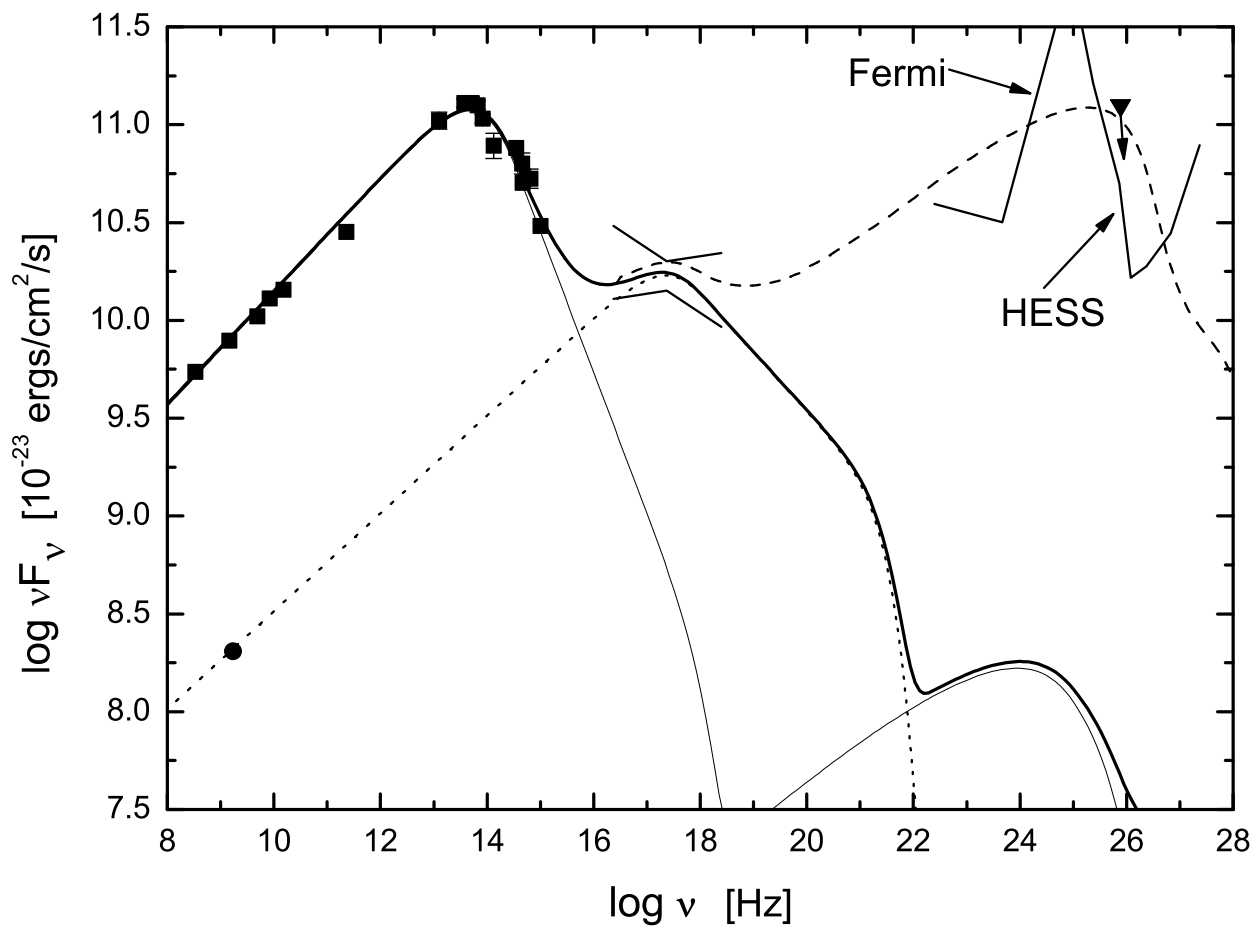} \caption{The data are same as Figure 1, except a
radio point (solid circle) is from Tingay et al. (2008). The thick
solid line is the total emission of the diffuse region (thin solid
line) and the sub-components (dotted line), given by the
synchrotron + SSC model under the equipartition magnetic field of
$B=4.0\times10^{-4}$ G, while the dashed line is the total
radiation in a small magnetic field strength $B=1.5\times10^{-5}$
G.
 \label{Fig:2}}
\end{figure}

\clearpage
\begin{figure}
\plotone{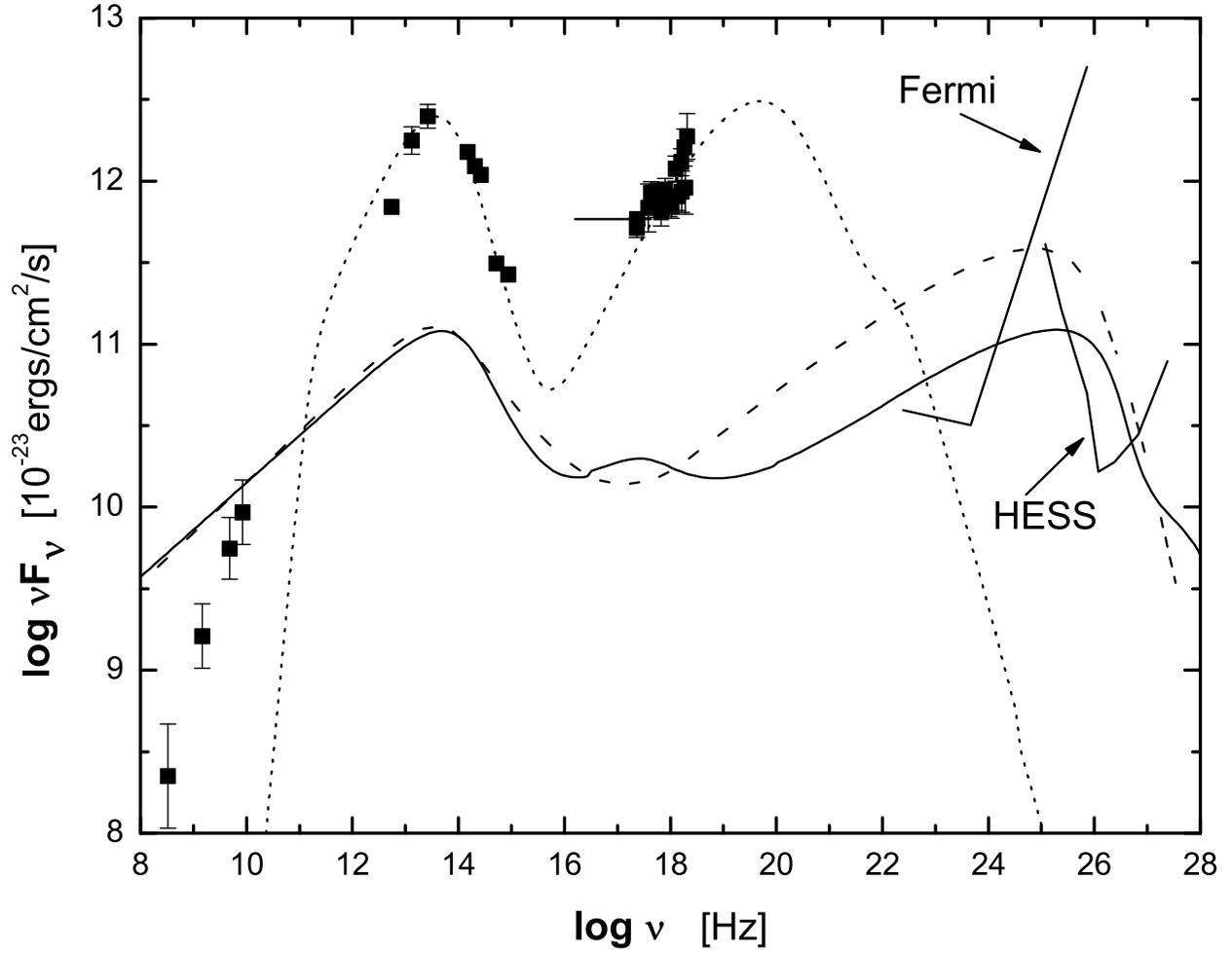} \caption{Comparison of SEDs for the nucleus
(squares/dotted line) and the WHS, represented by single-zone
model (dashed line) and multi-zone model (solid line),
respectively.
 \label{Fig:3}}
\end{figure}
\label{lastpage}

\end{document}